\documentclass[preprint,proceedings]{rmaa}

\usepackage{paralist}

\usepackage{psfrag,color}

\SetYear{2006}
\SetConfTitle{LARIM05}

\title{Galaxy Formation and Supernova Feedback} 

\author{
  P. B. Tissera,\altaffilmark{1,2} 
  C. Scannapieco,\altaffilmark{1,2}
  S. D. M. White\altaffilmark{3}
 and V. Springel \altaffilmark{3}}
\altaffiltext{1}{Instituto de Astronom\'\i{}a y F\'{\i}sica del Espacio (IAFE), Argentina.
patricia@iafe.uba.ar}
\altaffiltext{2}{Consejo Nacional de Investigaciones CienF\'{\i}ficas y T\'ecnicas (Conicet), Argentina.}
\altaffiltext{3}{Max-Planck Institute for Astrophysics, Germany.}

\shortauthor{Tissera et al.}
\shorttitle{Galaxy Formation and SN Feedback}

\listofauthors{Tissera, P. B., Scannapieco, C., White, S. D. M. & Springel, V.}
\indexauthor{Tissera, P. B.}
\indexauthor{Scannapieco, C.}
\indexauthor{White, S. D. M.}
\indexauthor{Springel, V.}

\abstract{We present a Supernova (SN) feedback model that succeeds at  describing  
the chemical and energetic effects of 
SN explosions in galaxy formation simulations. This new SN model has been coupled
to  GADGET-2 and works within a new multiphase scheme which allows the description 
of a co-spatial mixture of cold and hot interstellar medium phases. 
 No ad hoc scale-dependent  parameters are associated to these SN  and multiphase models making
them particularly suited to studies of galaxy formation in a cosmological framework.
Our SN model succeeds not only in setting a self-regulated star formation activity in galaxies
but in triggering collimated chemical-enriched galactic winds. The  effects of winds
vary with the virial mass of the systems so that the smaller the galaxy, the larger the fraction
of swept away gas and the stronger the decrease in its star formation activity. The fact that 
the fraction of ejected metals  exceeds 60 per cent regardless of mass, suggests that SN feedback
can be the responsible mechanism of  the enrichment of the intergalactic medium to the observed
levels. }

\addkeyword{Supernova}
\addkeyword{Galaxy Formation}
\addkeyword{Numerical Models}
\addkeyword{Cosmology}

\begin{document}
\maketitle

\section{Introduction}
\label{sec:intro}
The understanding of galaxy formation in hierarchical clustering scenarios remains
a primary goal for cosmology. Whereas gravitation governs  matter evolution in large scales,
  more complex
physical processes appear as principal actors at galaxy scales. 
Lagrangian-based models, such as Smoothed Particle Hydrodynamical (SPH) codes,
have been widely used in the last two decades to study galaxy formation.
 Several models have been developed
to describe gas cooling and star formation. However,
 problems arise owing to the fact that SPH tends to overestimate the densities of 
the hot gas, resulting in an overestimation of the cooling rates and an excessive
condensation of cold gas  (e.g. Springel \& Hernquist 2003). This excess of dense-cold gas
translates into high star formation rates.  Because
new stars tend to form in these dense-cold clumps with very short cooling times, if 
the energy released by SN explosions is directly thermalized within these regions,
no effects on the dynamical evolution of the
systems are achieved. Several strategies have been developed to overcome these
problems (e.g Marri \& White 2003 and references therein). 
In this work, we use previous ideas discussed
 by Marri \& White (2003) and develop a new numerical implementation for
SN feedback. We include a numerical treatment for both chemical and energy
release. No ad hoc numerical or scale-dependent parameters are needed in 
our SN model which makes it specially suited for cosmological studies.

\section{Analisis and Discussion}
\begin{figure}
    \begin{center}
      \includegraphics[width=7.5cm]{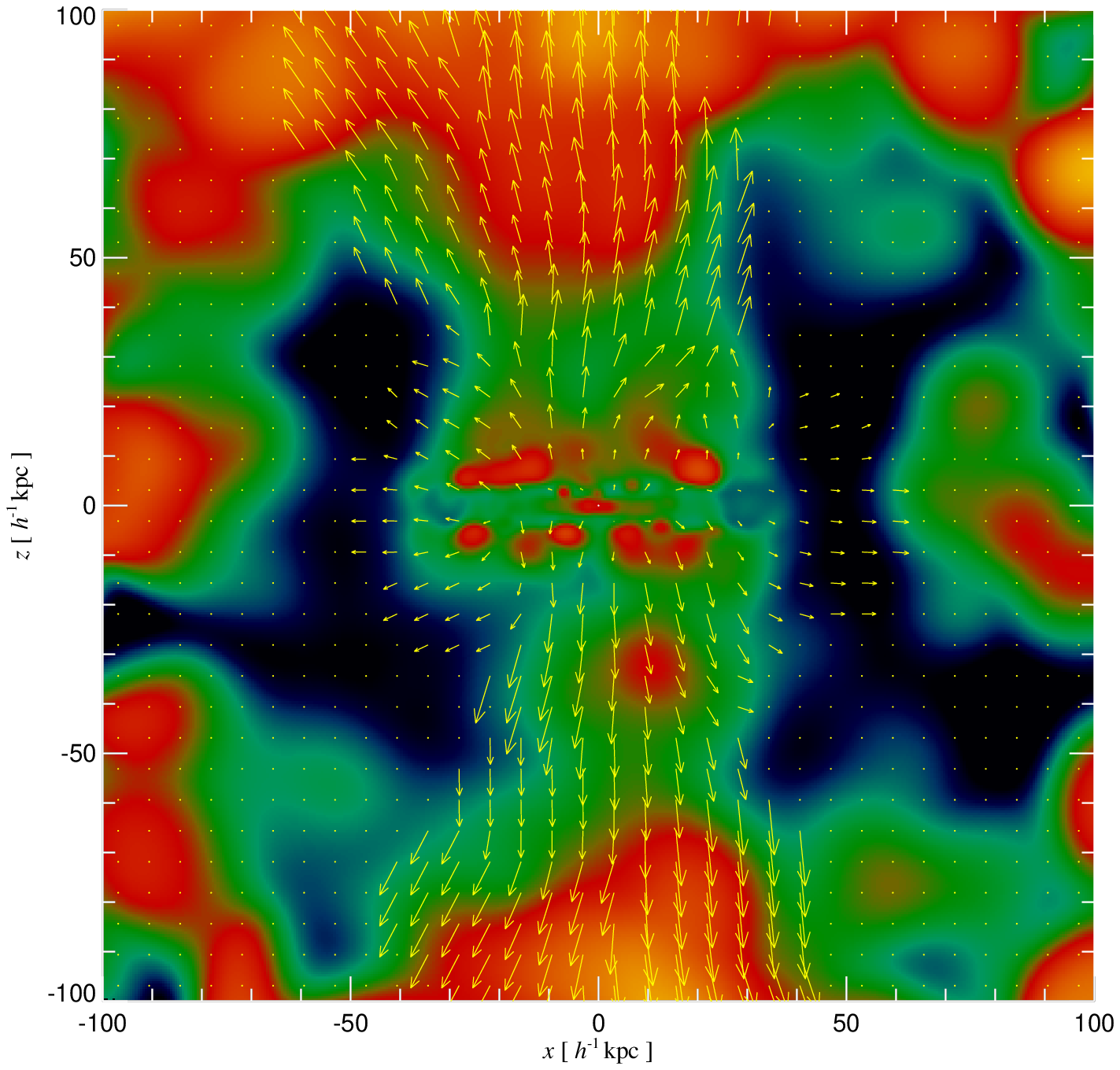}
      \includegraphics[width=7.5cm]{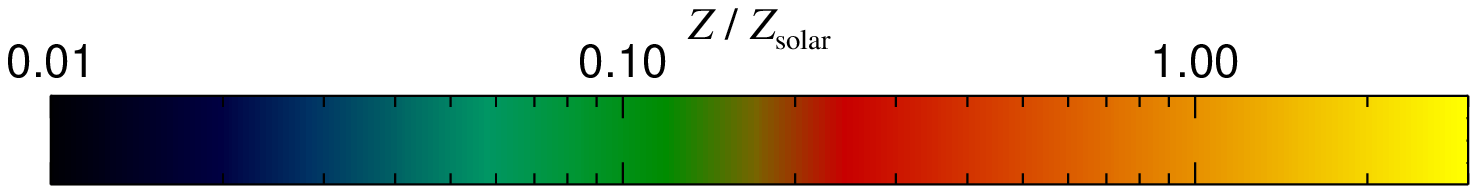}
    \end{center}
  \caption{{ Mean metallicity distribution of an edge-on disc of 
 $10^{12} M_{\odot} h^{-1}$ virial mass resolved initially with $10^4$ gas particles
and $10^4$ dark matter particles. The arrows point out toward the flow direction and
their lengths represent the mean velocity at the  corresponding point. The metallicity
scale is also shown.} \label{outflow} }
  \end{figure}

Our new SN feedback has been coupled to  GADGET-2 (Springel \& Hernquist 2002)
and is described in detail by Scannapieco et al. (2005, 2006a in preparation).
A new multiphase model which allows a better overlapping between gas clumps
of different densities and temperature has been developed. This multiphase model
prevents particles of very different entropy to be selected as neighbours. Hence,
diffuse, hot gas clumps can co-exist with cold and dense ones because
they are adequately decoupled within the SPH calculations. Both chemical
elements and energy released by SNII and SNIa are distributed within the gas
neighbours of stellar particles hosting the SN events. However, these neighbours are
segregated into cold and hot phases so that each of these phases receives
a given fraction of metals and energy. Whereas the hot gaseous phase stores the metals and
thermalizes instantaneously the energy, the cold phase gets chemically enriched but
the received energy is stored in a reservoir. Cold gas particles accumulate
energy by successive nearby SN events until the reservoir energy is enough to 
change the thermodynamical properties of the cold particle to match those of its hot environment. 
The properties of the hot environment of each cold gas particle are estimated from
those particles that do not consider it as a neighbour during SPH calculation.
This new SN model is capable of regulating the star formation activity
in systems of different virial masses 
in a self-consistent way. 
The success of this new SN model is owing to the fact that it works within a multiphase
scheme which defines the properties of each gas clump and its hot and cold environments
 on individual basis. 
It is a particle-particle criterium and there is no global definition involved.
Within this new SN scheme, collimated metal-enriched winds are consistently triggered. 
In Figure~\ref{outflow} we show the metallicity distribution for an edge-on isolated
disc galaxy test. Velocities  along the outflows can be up to 1000 ${\rm km s^{-1}}$
depending on how the SN energy is distributed among the hot and cold environments, and
on the potential well of the systems. Discs with less than  $10^{10} M_{\odot} h^{-1}$
are gas-depletted   more efficiently. Part of this gas is
not unbound and can be accreted later again producing new starburst episodes.

We used  this SN model in  high resolution cosmological simulations of Local group-type regions 
in a $\Lambda$-CDM scenario. We found
that this new SN model assures the conservation of the 
specific angular momentum of the disc component,
being able to produce extended exponential discs with
scale-lengths of $\approx 3$ kpc $h^{-1}$ ($h=0.7$). The stars in the bulge and halo components have
age and mean metallicity distributions which resembled those of the Milky-Way 
(see Scannapieco et al. 2006b, in preparation).

In summary, this new SN feedback model probes to be able to regulate the star formation
activity and to trigger enriched galactic winds, self-consistently. 
This new SN feedback model opens a new stage in the study of galaxy formation
and the history of chemical enrichment of the Universe.

\acknowledgements
PBT thanks the LOC and SOC of the LARIM05 for their support.
This work was partially supported by
Fundaci\'on Antorchas, Conicet, Prosul2005 and LENAC.

\end{document}